# The Orientation of Local Spatial Reference Frames in Quantum Teleportation

Douglas M. Snyder
Los Angeles, California

Abstract

Bennett et al.'s original article on quantum teleportation did not address explicitly synchronizing the orientation of the local spatial reference frames from and to which teleportaton occurs where the state teleported is spin 1/2 along the $z$ axis. They wrote that only two bits of information need to be conveyed classically where the state to be teleported concerns spin 1/2 along the $z$ axis. That only two bits of information need to be classically conveyed could imply that such synchronization is not needed for it would require more classically conveyed information. A gedankenexperiment is presented regarding the orientation of the local spatial reference frames exploring whether such synchronization is not needed even though it was apparently assumed by Bennett et al. that it was required. Physicists often *assume* that there is an absolute, as opposed to a relative, orientation of objects in space. Their assumption should be open to empirical test as any assumption in science should be, especially in view of evidence in psychology that the orientation of objects in space is not absolute.

Text

Quantum teleportation raises the question of the orientation of the local spatial reference frames from which and to which teleportaton occurs where the state teleported is spin 1/2 along the $z$ axis. The orientation of the local spatial reference frames might be important in quantum teleportation. Bennett and his colleagues [1] in their original article on quantum teleportation wrote that only two bits of information needed to be conveyed classically where the state to be teleported concerns spin 1/2 along the $z$ axis. These two bits account for the four possible results of the Bell-state measurement at the location where the quantum state is to be teleported from. It would be reasonable to think that Bennett and his colleagues thus indicated that essentially there need not be any prior agreement before quantum teleportation occurs between observers at the distant locations from where and to where teleportation occurs as regards orientation of the local spatial reference frames with one another. This possibility raises the question of if, and perhaps how, these local frames are related in terms of their orientation. Physicists often *assume* that there is an absolute, as opposed to a relative, orientation of objects in space, and this assumption apparently was held by Bennett et al. This assumption was apparently so firmly held by Bennett et al. that they did not explicitly note in their article the synchronization of the orientations of widely distant local reference frames from where and to where teleportation occurs that may be dependent on the absolute positioning of objects in space. Their assumption should be open to empirical test as any assumption in science should be, especially in view of evidence in psychology that the orientation of objects in space is not absolute. A gedankenexperiment is presented regarding the orientation of the local spatial reference frames exploring whether such synchronization is *not* needed.

Specifically, in the teleportation scheme proposed by Bennett et al. [1], a quantum state of a spin 1/2 particle X at one location can be teleported to another distant location where another spin 1/2 particle Y is located. An experimental realization following the basic idea of Bennett et al. has been accomplished [2].

Following Bennett et al., allow that in a gedankenexperiment the state to be teleported for a spin one-half particle X for the spin component along the $z$ axis is:

$$|\psi_X> = \alpha|\uparrow_X> + \beta|\downarrow_X> \quad (1)$$

After teleportation, perhaps including a rotation around the $x$, $y$, or $z$ axes where Y is located or perhaps allowing for an unimportant phase factor, the state of Y should be the same as that which characterized X before teleportation:

$$|\psi_Y> = \alpha|\uparrow_Y> + \beta|\downarrow_Y> \quad (2)$$

The following question arises: How does one determine the orientation of the $x$, $y$, or $z$ axes at Y which is distant from X, inasmuch as only two bits of information are classically communicated and Bennett et al. did not note any other information needed to be classically transmitted? When an observer obtains the classically communicated message regarding the nature of the Bell-state measurement at X, how does the observer at Y determine how the local, orthogonal spatial axes are located within which a rotation can be made, if needed, to achieve the same state for Y that characterized X before teleportation? What is the orientation frame within which these orthogonal axes for an observer at Y are located? Does one somehow attempt to define some

# The Orientation of

more general spatial coordinate frame in which the local spatial coordinate frames pertaining to X and Y are oriented, an attempt that would follow on an assumption that synchronization of the orientation of the local reference frames is necessary? Or is it possible that one does not need to be concerned about using a more general spatial reference frame with regard to the relative orientations of the local spatial coordinate frames for X and Y? In the latter case, the local coordinate frame would somehow be in alignment with each other in terms of local factors. What these factors are would require investigation.

To determine whether a more general spatial reference frame is relevant or whether the local spatial frames for X and Y are *not* located within a more general frame, our gedankenexperiment can be elaborated. One cannot simply accept without investigation that some prior coordination of orientations of the local reference frames for X and Y is not necessary, especially in lieu of the possibility of widely separated local reference frames in arbitrary orientations relative to one another [1]. In elaborating our gedankenexperiment, we will assume the position *not* commonly held that the orientations of the local spatial reference frames do not need to be synchronized. Consider a series of experimental runs (1,2,3…) where a number of particles $X_{1,2,3...}$ are produced in the exact same state (with the same wave function $\psi$) and teleported to $Y_{1,2,3...}$. Allow that $X_{1,2,3...}$ is prepared in the state (1) above in a lab located at one location on the earth's surface. Let $Y_{1,2,3...}$ be located in a lab such that $X_{1,2,3...}$ and $Y_{1,2,3...}$ are located at two ends of the earth's diameter and $Y_{1,2,3...}$ takes on state (2) after quantum teleportation. $X_{1,2,3...}$ and $Y_{1,2,3...}$ are at opposite locations on the earth's surface. How would an observer next to $Y_{1,2,3...}$ set up the lab so as to determine the *z* axis needed to interpret state (2) of $Y_{1,2,3...}$? Allow that the observers at $X_{1,2,3...}$ and $Y_{1,2,3...}$ do not know each others' relative orientations to one another. As before, consider the case where the Bell state measurement at $X_a$ and another particle yields a measurement indicating that state (1) is teleported with only a phase change to $Y_a$:

$$- |\psi_Y> = - (\alpha|\uparrow_Y> + \beta|\downarrow_Y>) \tag{3}$$

Allow further that the lab at $X_{1,2,3...}$ is set so that the *z* axis for $X_{1,2,3...}$ noted in state (1) is orthogonal to the earth's surface and that spin up along the *z* axis indicates the direction pointing away from the earth's surface. In a similar lab setup at $Y_{1,2,3...}$ where the *z* axis also points away from the earth's surface, the spin up direction for $Y_{1,2,3...}$ would be reversed. They would be in opposite directions when compared to a more general *z* axis that encompasses the local *z*

The Orientation of

axes for the labs where $X_{1,2,3...}$ and $Y_{1,2,3...}$ are located. An example of a more general *z* axis is depicted in Figure 1. If the results obtained for $Y_{1,2,3...}$ indicated that the state that had characterized $X_{1,2,3...}$ before teleportation has accurately teleported, then the data obtained in measurements on $Y_{1,2,3...}$ would

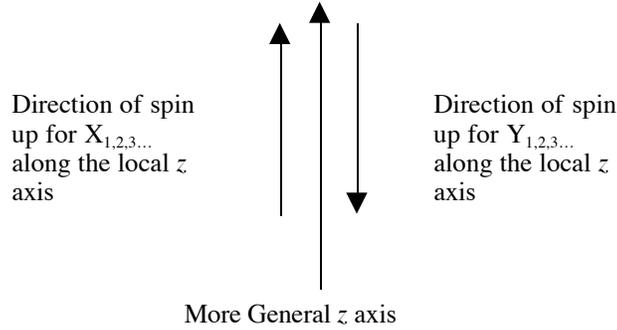

Direction of spin up for $X_{1,2,3...}$ along the local *z* axis

Direction of spin up for $Y_{1,2,3...}$ along the local *z* axis

More General *z* axis

Figure 1. Comparison of Directions for Spin Up for $X_{1,2,3...}$ and $Y_{1,2,3...}$ in Relation to a More General *z* axis

have been those obtained for $X_{1,2,3...}$ had they been made and the thesis that there is only a local nature for an orientation frame within which the local spatial coordinate frame is located would be supported (Table 1). These are the results predicted by Eqn. 2. If, on the other hand, there is a more general spatial frame within which the local ones for $X_{1,2,3...}$ and $Y_{1,2,3...}$ are located and which serves to orient the local *z* axes for $X_{1,2,3...}$ and $Y_{1,2,3....}$ regardless of how the observers set up their local spatial frames, the results obtained for $Y_{1,2,3...}$ would be different than those that would be obtained if the measurement had been made on $X_{1,2,3...}$ before teleportation. An example of these results, corresponding to the example in Figure 1, is displayed in Table 2. The results in Figure 2 are not those predicted by Eqn. 2 for $Y_{1,2,3...}$.

| Table 1 Percentage of Particles $Y_{1,2,3...}$ Expected to be Observed to Have Spin Up or Spin Down Along the Local *z*-axis Where Only Local Spatial Frames Exist ||
|---|---|
| ↑ | $|\alpha|^2$ |
| ↓ | $|\beta|^2$ |

- 4 -

The Orientation of

| Table 2 Percentage of Particles $Y_{1,2,3...}$ Expected to be Observed to Have Spin Up or Spin Down Along the Local $z$-axis Where a General Spatial Frame Exists Like that In Figure 1 and Supercedes Local Frames ||
|---|---|
| ↑ | $\|\beta\|^2$ |
| ↓ | $\|\alpha\|^2$ |

In more concrete terms, if for example $\alpha = 1/2$ and $\beta = \sqrt{(3/4)}$, if a local orientation frame is the only relevant frame, the percentage of particles $Y_{1,2,3...}$ with spin up is $|\alpha|^2 = 1/4$ and spin down is $|\beta|^2 = 3/4$. If a more general spatial frame with a $z$ axis like that in Figure 1 is relevant to the local spatial coordinate frames for $X_{1,2,3...}$ and $Y_{1,2,3...}$ and takes precedence, the percentage of particles $Y_{1,2,3...}$ with spin up is $|\beta|^2 = 3/4$ and spin down is $|\alpha|^2 = 1/4$. This result contradicts that expected by Eqn. 2.

The most general circumstances are that $Y_{1,2,3...}$ could be located in a local spatial reference frame with any orientation relative to that for $X_{1,2,3...}$. Bennett et al. [1] noted:

> Teleportation has the advantage of still being possible in situations where Alice and Bob, after sharing their EPR pairs, have wandered about independently and no longer know each others' locations. Alice cannot reliably send Bob the original quantum particle, or a spin-exchanged version of it, if she does not know where he is; but she can still teleport the quantum state to him, by broadcasting the classical information to all places where he might be.

Bennett et al. apparently maintained that the synchronization of Alice and Bob's spatial reference frames for teleportation is assumed in the above quote.

The general circumstances noted by Bennett et al. are the most informative as regards the concern raised in this paper. The general circumstances give rise to ample opportunities to test whether the orientation of the local reference frames need to be synchronized. In terms of our gedankenexperiment, the situation becomes more complicated because the relative orientations of the local spatial reference frames are arbitrary and all three orthogonal axes need to be taken into account.

If the hypothesis is supported that a general spatial reference frame is not relevant, then psychological factors should be investigated as factors in the local orientation frames within which the local spatial axes are placed. It has been shown that psychological factors are central to an individual's sense of the upright that would form the basis for the observer's personal spatial structure [e.g., 3,4,5,6,7,8]. Because of this feature of the person's sense of the upright, different retinal patterns may correspond to the same perception. In conjunction with the quantum mechanical principle that not more than one of a pair of canonically conjugate quantities can be known precisely at one time, these canonically conjugate quantities may both be registered on the retinas of observers' eyes (when one considers spin components of a spin 1/2 particle along orthogonal axes) while the visual perception of the individuals is essentially the same.

Also, this feature of perception in quantum mechanics supports the measurement of the $z$ axis of a spin 1/2 particle, for example, along any spatial axis the experimenter chooses, which experiment bears out. If no objective retinal pattern is associated in a one-to-one manner with a particular perception of spin along an axis (also indicating there is no absolute orientation of objects in space), then it is possible that the same retinal pattern may support spin component measurements along the same axis *even if the orientation of the axis varies in space*.[1] Specifically, the observer is determining the $z$ component of the spin 1/2 along an axis oriented in any rotated position in space.

It may be that the personal spatial structure of the observer serves as the foundation for the local orientation frame within which the spatial coordinate frame used to measure either X or Y locally if a more general spatial coordinate frame is not applicable. The possibility that the personal spatial structure of the observer is significant can be subjected to empirical test.

---

[1] It is possible that any idea from a major physical theory concerning measurement in the physical world has some root in the cognitive and/or perceptual capabilities of a person, even if it does not appear that these capabilities are directly related to the physical phenomena.